\documentclass[twocolumn,twoside,slac_two]{revtex4}
\usepackage{graphicx}
\usepackage{fancyhdr}
\begin{document}

\title{New dynamical gauges of the SM of EW interactions}

%

\author{B.A.Li}
\affiliation{Department of Physics and Astronomy, University of Kentucky, Lexington, KY 40506, USA}

\begin{abstract}
A new Lagrangian of EW interactions without spontaneous symmetry breaking, Higgs, and 
Fadeev-Popov procedure has been constructed. It consists of three parts: $SU(2)_L\times U(1)$ gauge fields, massive fermion
fields, and their interactions. In this theory the gauge coupling constants g, g', and fermion masses are the inputs.
A new mechanism of $SU(2)_L\times U(1)$ symmetry breaking caused by fermion masses has been
found. In the EW theory top quark mass plays a dominant role. 
This mechanism leads to nonperturbative generation of the gauge fixings and masses of Z and W fields. 
$m^2_W={1\over2}g^2 m^2_t,\;m^2_Z={1\over2}\bar{g}^2 m^2_t,\;G_F=\frac{1}{2\sqrt{2}m^2_t}$ are revealed from this theory.
The propagators of Z- and W-fields without quadratic divergent terms are derived.
Very heavy neutral and charged scalars are dynamically generated from this theory too.
\end{abstract}

\maketitle

\thispagestyle{fancy}



\section{New Lagrangian of EW interactions}
In the SM of electroweak interactions(EW) there are spontaneous symmetry breaking, Higgs, and Faddeev-Popov procedure.
The theory is successful in many aspects. However, there is no experimental indication for existence of Higgs.
Other theoretical approaches:
Supersymmetry models, Technichcolor, little Higgs, top quark condensate etc. have been proposed to modify the SM.
In this talk a new possible EW theory without introducing any unknown factors is discussed.
In both QED or QCD there are three parts: gauge fields, massive fermion fields, and interactions between them.
They are two very successful quantum field theories.
So far, three parts of the theory of EW interactions $SU(2)_L\times U(1)$ gauge fields, massive fermion fields, 
and the interactions between
$SU(2)_L\times U(1)$ gauge fields and fermions are indeed confirmed experimentally. Therefore, it is worth to 
study the Lagrangian of EW interactions without spontaneous symmetry breaking and Higgs~\cite{li1}
\begin{eqnarray} 
{\cal L}=
-{1\over4}A^{i}_{\mu\nu}A^{i\mu\nu}-{1\over4}B_{\mu\nu}B^{\mu\nu}
+\bar{q}\{i\gamma\cdot\partial-m_q\}q\nonumber
\end{eqnarray}
\begin{eqnarray}
+\bar{q}_{L}\{{g\over2}\tau_{i}
\gamma\cdot A^{i}+g'{Y\over2}\gamma\cdot B\}
q_{L}+\bar{q}_{R}g'{Y\over2}\gamma\cdot Bq_{R}\nonumber
\end{eqnarray}
\begin{eqnarray}
+\bar{l}\{i\gamma\cdot\partial-m_{l}\}l
+\bar{l}_{L}\{{g\over2}
\tau_{i}\gamma\cdot A^{i}+g'{Y\over2}\gamma\cdot B\}
l_{L}\nonumber
\end{eqnarray}
\begin{eqnarray}
+\bar{l}_{R}g'{Y\over2}\gamma\cdot B l_{R}.
\end{eqnarray}
The structure of this L is the same as QED and QCD. The difference is that the $SU(2)_L\times U(1)$ gauge symmetry is broken 
by fermion masses.
The masses of W and Z bosons cannot be added by hand. The term
\begin{equation}
\frac{1}{q^2-m^2}\frac{q_\mu q_\nu}{m^2}
\end{equation}
of the propagator of massive gauge boson causes problems for renormalization and unitarity.
The challenge for this theory is: generating the masses of gauge bosons while the terms(2) which causes quadratic 
divergence must be canceled.
\section{New mechanism of symmetry breaking}
The masses of gauge bosons violate the gauge symmetry. In a gauge field theory
they should be the effects of $SU(2)_L\times U(1)$ gauge symmetry breaking.
In Eq.(1) $SU(2)_L\times U(1)$ gauge symmetry is broken by the fermion masses.  
We need to study the relationship between the masses of gauge bosons and fermions.
It is found that the fermion masses lead to a new mechanism of chiral symmetry breaking.
This mechanism is inspired by Weinberg's second sum rule obtained by current algebra~\cite{wei}
\begin{equation}
m^2_a=2m^2_\rho.
\end{equation}
$a_1$ and $\rho$ mesons are chiral partners and they are degenerate under chiral symmetry.
Obviously, Weinberg's second sum rule implicates a new mechanism of chiral symmetry breaking.
Current algebra is not a field theory. 
Based on current algebra and QCD the Lagrangian of a chiral field theory of pseudoscalar, vector,
and axial-vector mesons has been constructed as(in the case of two flavor)~\cite{li2}
\begin{eqnarray}
{\cal L}=\bar{\psi}(x)(i\gamma\cdot\partial+\gamma\cdot v
+\gamma\cdot a\gamma_{5}-mu(x))\psi(x)\nonumber
\end{eqnarray}
\begin{eqnarray}
+{1\over 2}m^{2}_{0}(\rho^{\mu}_{i}\rho_{\mu i}+
\omega^{\mu}\omega_{\mu}+a^{\mu}_{i}a_{\mu i}+f^{\mu}f_{\mu}),
\end{eqnarray}
where \(u=e^{(i\pi+i\eta_0)\gamma_5}\), m is the constituent quark mass which originates in quark condensate.
In the limit, $m_q\rightarrow 0$, the Lagrangian(4) has global $U(2)_L\times U(2)_R$ symmetry 
and the $a_1$ and the $\rho$ fields
have the same mass. 
Integrating out the quark fields, the Lagrangian of the mesons is
derived, in which $a_1$ and $\rho$ are nonabelian gauge fields. This theory is phenomenologically successful.
The mass relation between the $a_1$ and the $\rho$ mesons is derived
\begin{equation}
(1-{1\over 2\pi^2 g^2})m^2_a=2m^2_\rho,
\end{equation}
where g is the universal coupling constant of this theory[3], which is determined by the decay rate of
$\rho\rightarrow ee^+$.
Eq.(5) modifies Weinberg's second sum rule. In Ref.[2] Weinberg made an assumption which is no longer required in 
this new theory[3].
This new formula fits the data better than Eq.(3). Obviously, Eq.(5) means a $U(2)_L\times U(2)_R$ chiral symmetry 
breaking revealed from this theory(4) which is $U(2)_L\times U(2)_R$ symmetric at tree level.
How the $a_1$ meson gains additional mass and where the additional factor
\((1-{1\over 2\pi^2 g^2})\) comes from?  
The $a_1$ field is coupled to the axial-vector current of the massive fermion. 
Because this current is not conserved the vacuum polarization of the $a_1$ field is expressed as
\begin{eqnarray}
\Pi_{\mu\nu}^{ij}(q^2)=\int d^4 x e^{iqx}\nonumber 
\end{eqnarray}
\begin{eqnarray}
<0|T\{\bar{\psi}(x)\gamma_\mu\gamma_5\tau_i\psi(x)
\bar{\psi}(0)\gamma_\mu\gamma_5\tau_j\psi(0)\}|0>\nonumber
\end{eqnarray}
\begin{eqnarray}
=\delta_{ij}\{F_1(q^2)(q_\mu q_\nu-q^2 g_{\mu\nu})+F_2(q^2)q_\mu q_\nu+{1\over2}\Delta m^2 g_{\mu\nu}\}.
\end{eqnarray}
$F_2$ and $\Delta m^2$ are the results of the nonconserved axial-vector current of the massive quark.
They vanish when $m\rightarrow0$. Because of nonconservation of the axial-vector current $F_1(q^2)$ 
is different from the one of the $\rho$ field. $F_1(q^2)$ is used to normalize the $a_1$ field and 
the additional factor is resulted in $F_1(q^2)$. 
\begin{equation}
\Delta m^2=6m^2,\;\;m^2_\rho=6m^2.
\end{equation}
Eq.(5) is obtained.
$F_2$ is finite and it is the gauge fixing of the nonabalian $a_1$ field.
Therefore, the gauge fixing and additional mass term of $a_1$ field are
dynamically generated from this theory.
In the Lagrangian(4) there are no kinetic terms for the meson fields, which
are generated from quark loop diagrams. The quark loop diagrams must be treated
nonperturbatively. According to Eq.(4), only one quark-loop diagrams are allowed to be
calculated. The gauge fixing and additional mass term of $a_1$ field are
nonperturbative solutions of this theory. 
This is the {\bf new mechanism of symmetry breaking}: nonconserved axial-vector current of massive quark
leads to nonperturbative generation of both new mass term and gauge fixing of the $a_1$ field.
In Ref.[3] the the vacuum polarization of the $a_1$ field has been calculated to the $4^{th}$ in covariant derivatives.
Besides the new mass form formula(5) all the three terms of Eq.(6) have been tested in studying
$a_1$-related meson physics. Without them PCAC cannot be satisfied in this theory. Therefore, the mechanism 
is supported by $a_1$-related meson physics. On the other hand, if ignoring 
the term ${1\over 2}m^{2}_{0}a^{\mu}_{i}a_{\mu i}$ of Eq.(3), the mass of the gauge boson is via this mechanism dynamically
generated by fermion mass, while the gauge fixing is
simultaneously generated by this new mechanism to cancel the term which causes quadratic divergence. Therefore,
this mechanism does solve the problem mentioned above. In this mechanism fermion mass is the origin of the mass 
of gauge boson and gauge fixing. It has been studied in Ref.[1] that the finite $F_2$ leads to a scalar field.

A charged vector current of massive fermions is not conserved too. The vacuum polarization of a charged vector
current of massive fermions takes the same expression as Eq.(6). Besides $F_1$ there are $F_2$ and $\Delta m^2$
too and they vanish when the masses of fermions go to zero[1]. Therefore, nonconserved charged vector current 
of massive fermions
leads to nonperturbative generation of both new mass term and gauge fixing of the charged vector fields.
In Ref.[1] a model has been constructed to show this mechanism
\begin{eqnarray}
{\cal L}=i\bar{u}\gamma\cdot\partial u +i\bar{d}\gamma\cdot\partial d
+\bar{u}\gamma_\mu dv^{-}_\mu+\bar{d}\gamma_\mu uv^{+}_\mu\nonumber
\end{eqnarray}
\begin{eqnarray}
-m_u\bar{u}u-m_d\bar{d}d.
\end{eqnarray}
The amplitude of the vacuum polarization is derived
\begin{eqnarray}
\Pi^{v}_{\mu\nu}={1\over2}(q_\mu q_\nu-p^2 g_{\mu\nu})F_{v1}(q^2)
+F_{v2}(q^2)q_\mu q_\nu\nonumber
\end{eqnarray}
\begin{eqnarray}
+{1\over2}m^2_{v}g_{\mu\nu}.
\end{eqnarray}
\(m_u\neq m_d\) leads to $F_{v2}(q^2)$ and $m^2_{v}$.
The expressions of $F_{v1}(q^2)$, $F_{v2}(q^2)$, and $m^2_{v}$ can be found in Ref.[1].

In The Lagrangian(1) the Z-field is coupled to axial-vector currents of massive fermions and W-fields are coupled
to charged vector currents of massive fermions. The new mechanism of symmetry breaking can be applied to this EW theory.
In Eq.(1) there are no mass terms for the gauge fields and the gague symmetries are broken by the fermion masses, 
the propagators of gauge bosons cannot be defined and perturbation theory related to gauge fields 
cannot be defined at the tree level. 
In order to construct the perturbation theory the vacuum polarizations of 
Z- and W-fields must be treated nonperturbatively. The masses and gauge fixings of Z and W fields are via this new 
mechanism dynamically generated from this theory. They are nonperturbative solutions of Eq.(1).
As shown above, fermion mass plays essential role in this new mechanism. 
There is very special effect in the EW theory: the heavy top quark of the EW theory(1)
provides strong strength for the symmetry breaking.  
\section{$m_Z$, propagator of Z-field, neutral scalar}
The couplings between Z-field and fermions are found from Eq.(1).
We start from the t- and b-quark generation. The coupling 
between Z-boson and t and b quarks is found
\begin{eqnarray}
{\cal L}={\bar{g}\over4}\{
(1-{8\over3}\alpha)\bar{t}\gamma_\mu t+
\bar{t}\gamma_\mu\gamma_5  t\}Z^\mu\nonumber
\end{eqnarray}
\begin{eqnarray}
-{\bar{g}\over4}\{(1-{4\over3}\alpha)\bar{b}\gamma_\mu b
+\bar{b}\gamma_\mu\gamma_5 b\}Z^\mu,
\end{eqnarray}
where \(\alpha=sin^2 \theta_W \). In Eq.(10) there are axial-vector currents of
massive t- and b- quarks. Other generations of quarks and leptons can be added.
The vacuum polarization of Z-field is expressed as
\begin{equation}
\Pi^Z_{\mu\nu}={1\over2}F_{Z1}(z)(p_\mu p_\nu-p^2 g_{\mu\nu})+F_{Z2}(z)
p_\mu p_\nu+{1\over2}m^2_Z g_{\mu\nu},
\end{equation}
\begin{eqnarray}
F_{Z1}=1+\frac{\bar{g}^2}{64\pi^2}\{\frac{D}{12}\Gamma(2-{D\over2})
[N_C y_q\sum_q(\frac{\mu^2}{m^2_q})^{{\epsilon\over2}}\nonumber
\end{eqnarray}
\begin{eqnarray}
+y_l\sum_l(\frac{\mu^2}{m^2_l})^{{\epsilon\over2}}]\nonumber 
\end{eqnarray}
\begin{eqnarray}
-2[N_C y_q\sum_q f_1(z_q)+y_l\sum_lf_1(z_l)]\nonumber
\end{eqnarray}
\begin{eqnarray}
+2[\sum_q f_2(z_q)
+\sum_{l=e,\mu,\tau}
f_2(z_l)]\},\nonumber 
\end{eqnarray}
\begin{eqnarray}
F_{Z2}=-\frac{\bar{g}^2}{64\pi^2}\{N_C\sum_qf_2(z_q)
+\sum_{l=e,\mu,\tau}f_2(z_l)\},\nonumber
\end{eqnarray}
\begin{eqnarray}
m^2_Z={1\over8}{\bar{g}^2\over(4\pi)^2}D\Gamma(2-{D\over2})
\{N_c\sum_{q}m^2_q
({\mu^2\over m^2_q})^{{\epsilon\over2}}\nonumber
\end{eqnarray}
\begin{eqnarray}
+\sum_l m^2_l({\mu^2\over m^2_l})^
{{\epsilon\over2}}\}.
\end{eqnarray}
where \(y_q=1+(1-{8\over3}\alpha)^2\) for \(q=t,c,u\),
\(y_q=1+(1-{4\over3}\alpha)^2\)
for
\(q=b,s,d\), \(y_l=1+(1-4\alpha)^2\), for \(l=\tau, \mu, e\), \(y_l=2\) for
\(l=\nu_e,\nu_\mu,\tau_\mu\),
\(z_i={p^2\over m^2_i}\) and the functions $f_{1,2}$
are defined 
\begin{eqnarray}
f_1(z)=\int^1_0 dx x(1-x)log\{1-x(1-x)z\},\nonumber
\end{eqnarray}
\begin{eqnarray}
f_2(z)={1\over z}\int^1_0 dx log\{1-x(1-x)z\}.
\end{eqnarray}
Both the vector and axial-vector couplings contribute to $F_{Z1}$, which is used to renormalize the Z-field.

$m^2_Z$ is resulted in the axial-vector current of massive fermions. If $m_f\rightarrow 0$, 
$m^2_Z$ goes to zero too. The fermion masses are the origin of the mass of Z-boson.
In Eq.(11) there are divergences and renormalization are required. In the Lagrangian(1) there is no mass term of Z-field 
which, \({1\over2}m^2_Z Z_\mu Z_\mu\), is dynamically generated. It is better to do the renormalization of $m^2_Z$.
Eq.(12) shows $m^2_Z\propto m^2_f$ and top quark mass is dominant. Ignoring all other fermion masses, it is obtained
\begin{equation}
m^2_Z(physical)={N_C\over(4\pi)^2}{D\over4}\Gamma(2-{D\over2})
({\mu^2\over m^2_t})^{{\epsilon\over2}}m^2_Z,
\end{equation}
\begin{equation}
m^2_Z(physical)={1\over2}\bar{g}^2 m^2_t.
\end{equation}
This formula agrees with data very well. 

Only the axial-vector coupling contributes to $F_{Z2}$. $F_{Z2}$
is finite and negative. $F_{Z2}$ vanishes when the fermion masses are taken to be zero.
$F_{Z2}$ means a new scalar field $\phi^0$[1].
\begin{equation}
F_{Z2}(z)=\xi_Z+(p^2-m^2_{\phi^0})G_{Z2}(p^2),
\end{equation}
where $m^2_{\phi^0}$ is the mass of the new neutral spin-0 field,
$G_{Z2}$ is the radiative correction of this term, and $\xi_Z$ is the gauge fixing parameter
\begin{equation}
\xi_Z
=F_{Z2}|_{p^2=m^2_{\phi^0}}.
\end{equation}
The new free Lagrangian of the Z-field
is constructed as
\begin{equation}
{\cal L}_{Z0}=-{1\over4}(\partial_\mu Z_\nu-\partial_\nu Z_\mu)^2+\xi_Z(\partial_\mu Z^\mu)^2+{1\over2}m^2_Z Z^2_\mu.
\end{equation}
It is a Stueckelberg's Lagrangian.  
$\partial_\mu Z^\mu$ is a scalar field
\begin{eqnarray}
\partial^2(\partial_\mu Z^\mu)-{m^2_Z\over2\xi_Z}(\partial_\mu Z^\mu)=0,\nonumber
\end{eqnarray}
\begin{eqnarray}
Z_\mu=Z'_\mu\pm{1\over m_Z}\partial_\mu\phi^{0},\nonumber
\end{eqnarray}
\begin{eqnarray}
\partial_\mu Z'^\mu=0,\nonumber
\end{eqnarray}
\begin{eqnarray}
\phi^0=\mp{m_Z\over m^2_{\phi^0}}\partial_\mu Z^\mu,\nonumber
\end{eqnarray}
\begin{eqnarray}
\partial^2\phi^0-{m^2_Z\over2\xi_Z}\phi^0=0.
\end{eqnarray}
The mass of $\phi^0$ is the solution of the equation
\begin{eqnarray}
2m^2_{\phi^0}F_{Z2}|_{p^2=m^2_{\phi^0}}+m^2_Z=0,\nonumber
\end{eqnarray}
\begin{eqnarray}
m^2_{\phi^0}=-{m^2_Z\over2\xi_{Z}}.
\end{eqnarray}
The equation is expressed as
\begin{eqnarray}
3\sum_q{m^2_q\over m^2_Z}z_q f_2(z_q)+\sum_l {m^2_l\over m^2_Z}z_l
f_2(z_l)={32\pi^2\over\bar{g}^2}.
\end{eqnarray}
For $z>4$ it is found
\begin{eqnarray}
f_2(z)=-{2\over z}-{1\over z}
(1-{4\over z})^{{1\over2}}log\frac{1-(1-{4\over z})^{{1\over2}}}
{1+(1-{4\over z})^{{1\over2}}}.
\end{eqnarray}
Top quark dominates this equation
and it has a solution at very large value of z
\begin{eqnarray}
\frac{2(4\pi)^2}{\bar{g}^2}+\frac{6m^2_t}{m^2_Z}=3{m^2_t\over m^2_Z}log
{m^2_{\phi^0}\over m^2_t},\nonumber
\end{eqnarray}
\begin{eqnarray}
m_{\phi^0}=m_t e^{\frac{m^2_z}{m^2_t}{16\pi^2\over3\bar{g}^2}+1}
=m_t e^{28.4}=3.78\times10^{14}GeV,
\end{eqnarray}
The dynamically generated scalar is very heavy. 
\begin{eqnarray}
\xi_Z=-1.18\times10^{-25}.
\end{eqnarray}
The propagator of Z boson is found
\begin{eqnarray}
\Delta_{\mu\nu}=
\frac{1}{q^2-m^2_Z}\{-g_{\mu\nu}+(1+\frac{1}{2\xi_Z})\frac{q_\mu q_\nu}{
q^2-m^2_{\phi^0}}\},
\end{eqnarray}
It can be separated into two parts
\begin{eqnarray}
\Delta_{\mu\nu}=
\frac{1}{p^2-m^2_Z}\{-g_{\mu\nu}+\frac{p_\mu p_\nu}{
m^2_Z}\}\nonumber
\end{eqnarray}
\begin{eqnarray}
-\frac{1}{m^2_Z}\frac{q_\mu p_\nu}{q^2
-m^2_{\phi^0}}.
\end{eqnarray}
The first part is the propagator of the physical spin-1 Z boson and
the second part is the propagator of a new
neutral spin-0 meson, $\phi^0$. The minus sign of the scalar part indicates that the scalar has
negative metric.

The couplings between $\partial_\mu \phi^0$ and
fermions are found
\begin{eqnarray}
\lefteqn{{\cal L}=\pm{1\over m_Z}{\bar{g}\over4}\{
(1-{8\over3}\alpha)\bar{t}\gamma_\mu t+
\bar{t}\gamma_\mu\gamma_5  t\}\partial_\mu\phi^0}\nonumber \\
&&\pm{1\over m_Z}{\bar{g}\over4}\{-(1-{4\over3}\alpha)\bar{b}\gamma_\mu b
-\bar{b}\gamma_\mu\gamma_5 b\}\partial_\mu\phi^0\nonumber \\
&&\pm{1\over m_Z}{\bar{g}\over4}\{
\bar{\nu_e}\gamma_\mu(1+\gamma_5) \nu_e
-(1-4\alpha)\bar{e}\gamma_\mu e\nonumber \\
&&-\bar{e}\gamma_\mu\gamma_5 e\}\partial_\mu \phi^0.
\end{eqnarray}
The couplings with other generations of fermions take the same form.
Using the equations of fermions, it can be found that
the couplings between $\phi^0$ and fermions are
\begin{equation}
\pm{1\over m_Z}{\bar{g}\over4}2i\sum_i m_i\bar{\psi}_i\gamma_5\psi_i \phi^0,
\end{equation}
where i stands for the type of fermion.
Eq.(28) shows that the coupling between
$\phi^0$ and fermion is proportional to the mass of the fermion. 
The couplings between $\phi^0$ and
intermediate bosons can be obtained too.
It is necessary to point out that because ther propagators of the gauge
fields cannot be defined from Eq.(1), only single fermion-loops are 
allowed to be calculated.
\section{$m_W$, propagator of W-field, charged scalars}
The couplings between W-fields and fermions are found from Eq.(1)
\begin{equation}
{\cal L}={g\over4}\bar{\psi}\gamma_\mu(1+\gamma_5)\tau^i\psi W^{i\mu},
\end{equation}
Charged axial-vector and vector currents of massive fermions are shown in eq.(29).
Therefore, the new mechanism of symmetry breaking can be applied.
The expression of the vacuum polarization of fermions is obtained
\begin{equation}
\Pi^W_{\mu\nu}=F_{W1}(p^2)(p_\mu p_\nu-p^2 g_{\mu\nu})+2F_{W2}(p^2)
p_\mu p_\nu+m^2_W
g_{\mu\nu},
\end{equation}
where
\begin{eqnarray}
\lefteqn{F_{W1}(p^2)=1+{g^2\over32\pi^2}D\Gamma(2-{D\over2})\int^1_0 dx
x(1-x)}\nonumber \\
&&\{N_C\sum_{iq}({\mu^2\over L^i_q})^{{\epsilon\over2}}
+\sum_{il}({\mu^2\over L^i_l})^{{\epsilon\over2}}\}\nonumber \\
&&-{g^2\over16\pi^2}\{N_C\sum_{iq} f^i_{1q}+\sum_{il}
f^i_{1l}\}\nonumber \\
&&+{g^2\over16\pi^2}\{N_C\sum_{iq} f^i_{2q}+\sum_{il}f_{2l}\},\\
&&F_{W2}(p^2)=-{g^2\over32\pi^2}\{N_C\sum_{iq} f^i_{2q}+\sum_{il}
f^i_{2l}\},\\
&&
m^2_W={g^2\over4}{1\over(4\pi)^2}D\Gamma(2-{D\over2})
\int^1_0 dx\{N_c\sum_{iq}L^i_q
({\mu^2\over L^i_q})^{{\epsilon\over2}}\nonumber \\
&&+\sum_{il}L^i_l({\mu^2\over L^i_l})^
{{\epsilon\over2}}\}.
\end{eqnarray}
where
\begin{eqnarray}
L^1_q =m^2_b x+m^2_t (1-x),\nonumber
\end{eqnarray}
\begin{eqnarray}
L^2_q =m^2_s x+m^2_c (1-x),\nonumber
\end{eqnarray}
\begin{eqnarray}
L^3_q =m^2_d x+m^2_u (1-x),
\end{eqnarray}
\[L^1_l =m^2_e x,\]
\[L^2_l =m^2_\mu x,\]
\[L^3_l =m^2_\tau x,\]
\begin{eqnarray}
\lefteqn{
f^i_{1q}=\int^1_0 dx x(1-x)log[1-x(1-x){p^2\over L^i_q}]},\\
&&f^i_{1l}=\int^1_0 dx x(1-x)log[1-x(1-x){p^2\over L^i_l}],\\
&&f^i_{2q}={1\over p^2}\int^1_0 dx L^i_q log[1-x(1-x)
{p^2\over L^i_q}],\\
&&f^i_{2l}={1\over p^2}\int^1_0 dx L^i_l log[1-x(1-x){p^2\over L^i_l}].
\end{eqnarray}
The function $F_{W1}(p^2)$ is used to renormalize the W-fields.

$m^2_W$ is via the new mechanism originated in fermion masses. The mass term,
$m^2_W W^+_\mu W^{-\mu}$, is dynamically generated. It is required to do mass
renormalization of W-fields. Once again, top quark mass is dominant in $m^2_W$.
Ignoring all other fermion masses, it is obtained
\begin{equation}
m^2_W(physical)={N_C\over(4\pi)^2}{D\over4}\Gamma(2-{D\over2})
({\mu^2\over m^2_t})^{{\epsilon\over2}}m^2_W,
\end{equation}
\begin{equation}
m^2_W(physical)={1\over2}g^2 m^2_t.
\end{equation}
This formula agrees with data very well. 

Using Eqs.(15,40), it is obtained
\begin{equation}
\frac{m^2_W}{m^2_Z}=\frac{g^2}{\bar{g}^2}=cos^2\theta_W.
\end{equation}
\begin{equation}
G_F=\frac{1}{2\sqrt{2}m^2_t}.
\end{equation}
The study of the correction of Eqs.(15,40,41,42) can be done.

The $F_{W2}$ is finite. It is via the new mechanism obtained.
When $m_f\rightarrow 0$, it goes to zero. $F_{W2}$ is negative and 
leads to the existence of two charged spin-0 states, $\phi^\pm$.
$F_{W2}$ is rewritten as
\begin{eqnarray}
\lefteqn{F_{W2}=\xi_W+(p^2-m^2_{\phi_W})G_{W2}(p^2),}\\
&&\xi_W=F_{W2}(p^2)|_{p^2=m^2_{\phi_W}},
\end{eqnarray}
where $G_{W2}$ is the radiative correction of the term
$(\partial_\mu W^\mu)^2$ and $m^2_{\phi_W}$ is the mass of the charged spin-0
states, $\phi^\pm$. 
The free part of the Lagrangian of W-field
is redefined as
\begin{eqnarray}
{\cal L}_{W0}=-{1\over2}(\partial_\mu W^+\nu-\partial_\nu W^+_\mu)
(\partial_\mu W^-_\nu-\partial_\nu W^-_\mu)\nonumber
\end{eqnarray}
\begin{eqnarray}
+2\xi_W\partial_\mu W^{+^\mu}
\partial_\nu W^{-^\nu}
+m^2_W W^+_\mu W^{-\mu}.
\end{eqnarray}
It is a Stueckelberg's Lagrangian.

The divergences of the W-fields are independent degrees of freedom
\begin{equation}
\partial^2(\partial_\mu W^{\pm\mu})-{m^2_W\over2\xi_W}(\partial_\mu
W^{\pm\mu})=0.
\end{equation}
$\partial_\mu W^{\pm\mu}$ are spin-0 fields. Therefore,
the W field of the SM has four independent
components.
The W-field is decomposed as
\begin{eqnarray}
\lefteqn{W^\pm_\mu=W'^\pm_\mu \pm{1\over m_W}\partial_\mu \phi^\pm,}\\
&&\partial\mu W'^{\pm\mu}=0,\\
&&\phi^\pm=\mp{m_W\over m^2_{\phi_W}}\partial_\mu W^{\pm\mu},
\end{eqnarray}
where $W'$ is a massive spin-1 field and $\phi^{\pm}$ are spin-0 fields.
The equation of $\phi^\pm$ is derived from Eqs.(46)
\begin{equation}
\partial^2\phi^\pm-{m^2_W\over2\xi_{W}}\phi^\pm=0.
\end{equation}
The
mass of $\phi^\pm$ is determined by the equation
\begin{equation}
2m^2_{\phi_W}F_{W2}(p^2)|_{p^2=m^2_{\phi_W}}+m^2_W=0
\end{equation}
and
\begin{equation}
m^2_{\phi_W}=-{m^2_W\over 2\xi_{W}}.
\end{equation}
Numerical calculation shows that top quark is dominant in $F_{W2}$. Keeping the
contribution of top quark only, Eq.(51) becomes
\begin{equation}
{p^2\over m^2_W}F_{W2}=-\frac{3g^2}{32\pi^2}{m^2_t\over m^2_W}\{-{3\over4}
+{1\over2z}
+[{1\over2}-{1\over z}+{1\over2z^2}]log(z-1)\},
\end{equation}
where \(z={p^2\over m^2_t}\).
Eq.(53) has a solution at very large z. At very large z
we have
\begin{equation}
{p^2\over m^2_W}F_{W2}=-{3g^2\over64\pi^2}{m^2_t\over m^2_W}logz.
\end{equation}
The mass of $\phi^\pm$ is determined to be
\begin{equation}
m_{\phi_W}=m_t e^{{16\pi^2\over3g^2}{m^2_W\over m^2_t}}=m_t e^{27}
=9.31\times10^{13}GeV,
\end{equation}
and
\[\xi_W=-3.73\times10^{-25}.\]
The charged $\phi^\pm$ are very heavy too.

The propagator of W-field is derived from Eq.(45)
\begin{equation}
\Delta^W_{\mu\nu}=
\frac{1}{p^2-m^2_W}\{-g_{\mu\nu}+(1+\frac{1}{2\xi_W})\frac{p_\mu p_\nu}{
p^2-m^2_{\phi_W}}\},
\end{equation}
and it can be separated into two parts
\begin{equation}
\Delta^W_{\mu\nu}=
\frac{1}{p^2-m^2_W}\{-g_{\mu\nu}+\frac{p_\mu p_\nu}{
m^2_W}\}-\frac{1}{m^2_W}\frac{p_\mu p_\nu}{p^2
-m^2_{\phi_W}}.
\end{equation}
The first part of Eq.(57) is the propagator of physical
spin-1 W-field and the second
part is related to the propagator of the $\phi^\pm$ field. 

As shown above, the gauge fixings $\xi_z$ and $\xi_W$ are dynamically generated and
their values are fixed. There are no other ghosts associated with these gauge fixings.
Therefore, this scheme is completely different with Fadeev-Popov procedure of the SM.

The Lagrangian of interactions between fermions and $\partial_\mu \phi^\pm$
field is
found from Eqs.(29,47)
\begin{equation}
{\cal L}_{q\phi}=\pm{1\over m_W}{g\over4}\sum_j\bar{\psi}_j
\gamma_\mu(1+\gamma_5)
\tau^i\psi_j
\partial^\mu\phi^{i},
\end{equation}
where j is the type of the fermion and
\begin{equation}
\phi^1={1\over\sqrt{2}}(\phi^+ +\phi^-)\;\;\;
\phi^2={1\over \sqrt{2}i}(\phi^+ -\phi^-).
\end{equation}
Using the dynamical equations of fermions of the SM, for t and b quark
generation we obtain
\begin{equation}
{\cal L}'_{q\phi}=\pm{i\over m_W}{g\over4}(m_t+m_b)
\{\bar{\psi}_t\gamma_5\psi_b\phi^+
+\bar{\psi}_b\gamma_5\psi_t\phi^-\}.
\end{equation}
The coupling is proportional to the fermion mass. The couplings between other
generations of
fermion and $\phi^\pm$ field are similar to Eq.(60).

It is interesting to notice that coupling between the charged scalars and the photon takes the form
\begin{equation}
{\cal L}={4ie\over m^2_W}F^{\mu\nu}\partial_\mu\phi^+\partial_\nu\phi^-.
\end{equation}
The strength of the electromagnetic interactions is at the order of weak interactions.
\section{perturbation theory}
Using Eqs.(18,45) and the free Lagrangian of fermions(1), the perturbation theory of this new EW theory
can be constructed.
\section{Conclusions}
In the new theory of EW interactions(1) the gauge couplings g, g', and fermion masses are the parameters
of this theory. A new mechanism of symmetry breaking by fermion masses has been found. This new mechanism is 
supported by meson physics. In EW theory(1) top quark mass plays dominant role in this mechanism.
Gauge fixings and masses of Z and W bosons are via this new mechanism dynamically generated. They are
nonperturbative solutions of this theory. New propagators of Z- and W-fields are derived, in which
there are no quadratic divergent terms. A neutral and charged scalar fields are nonperturbative solutions
of this EW theory. These scalar fields have negative metric. Perturbation theory is constructed. 
It is worth to point out that the gauge fixings and heavy scalar fields 
are still nonperturbative solutions of the SM of EW interactions with spontaneous symmetry breaking and Higgs. 
In the new perturbation of theory of the SM
Fadeev-Popov procedure is no longer required. The question can be asked is that 
from the new mechanism the scalar bosons can be dynamically 
generated why we need to add Higgs to theory by hand.






\end{document}